\documentclass[]{raa}          

\usepackage{graphicx,times}           
\usepackage{natbib}
\usepackage{longtable}
\newcommand{\atlas}{{ATLAS$^{\rm 3D}$}}

\begin{document}

   \title{Revisiting the Dichotomy of Early-type Galaxies
}

   \volnopage{Vol.0 (200x) No.0, 000--000}   
   \setcounter{page}{1}       

   \author{Yan-Qin He
      \inst{1,2,3}
   \and Cai-Na Hao
      \inst{3}
   \and Xiao-Yang Xia
      \inst{3}
   }

   \institute{
University of Chinese Academy of Sciences, Beijing 100049, China
       \and
National Astronomical Observatories, Chinese Academy of Sciences, 20A Datun Road, Chaoyang District, Beijing 100012, China
        \and
Tianjin Astrophysics Center, Tianjin Normal University, Tianjin 300387, China \\{\it cainahao@gmail.com }
   }

   \date{Received~~ ; accepted~~ }

\abstract{We study the relationship among isophotal shapes, central light profiles and kinematic properties
of early-type galaxies (ETGs) based on a compiled sample of 184 ETGs.
These sample galaxies are included in the Data Release 8 of Sloan Digital Sky Survey (SDSS DR8)
and have central light profiles and kinematic properties available from the literature,
which were measured based on Hubble Space Telescope ({\it HST}) and \atlas\
integral-field spectrograph (IFS) observations, respectively.
We find that there is only a weak correlation between the isophotal shape ($a_{4}/a$) and
the central light profile (within $1\,\rm kpc$) of ETGs. About two-fifths of ``core''
galaxies have disky isophotes, while one-third of ``power-law'' galaxies are boxy deviated.
Our statistical results also show that there are weak correlations between
galaxy luminosity and dynamical mass with $a_{4}/a$, but such correlations are
tighter with central light profile. Moreover, no clear link has been found
between the isophotal shape and the S\' ersic index. Comparisons show that
there are similar correlations between $a_{4}/a$ and ellipticity and between $a_{4}/a$ and
specific angular momentum $\lambda_{R_e/2}$ for ``power-law'' ETGs, but there are no such correlations
for ``core'' ETGs.
Therefore, we speculate that the bimodal classifications for ETGs are not as simple as
previously thought, 
though we also find that the most deviated disky ETGs are ``power-law'', more elongated and fast rotators.
}

   \authorrunning{He et al. }            
   \titlerunning{Revisiting the Dichotomy of Early-type Galaxies}  

   \maketitle
%________________________________________________ sections below
%
\section{Introduction}           
\label{sect:intro}

The formation or assembly history of early-type galaxies (hereafter ETGs) has
been a hot topic in the field of galaxy formation and evolution. Since the
formation history of ETGs can imprint on their photometric and kinematical
properties, extensive efforts have been made to explore these properties
utilizing both imaging and spectroscopic observations.  Since then significant
progress has been made. Especially, dichotomies in isophotal shapes, nuclear
light profiles and kinematics were found. In most cases, isophotal shapes are
not perfect ellipses. Fourier analyses of the deviations from ellipses showed
that the most significant non-zero component is the coefficient of the fourth
cosine term (i.e., $a_{4}/a$; see Lauer 1985; Bender et al. 1988, 1989; Hao et
al. 2006; Kormendy et al. 2009b).  The sign of $a_{4}/a$ was used to divide ETGs
into two classes: boxy ($a_{4}/a<0$) and disky ($a_{4}/a>0$) (Bender et al.
1988; Faber et al. 1997).  Interestingly, some other properties also show that
boxy and disky ETGs are two different populations.  Boxy ETGs tend to be
bright, have strong radio and X-ray emission, and rotate slowly while disky
ETGs are faint, radio-quiet, have no X-ray hot gaseous halos and show regular
rotation patterns.

Similarly, dichotomy was also found in the central properties of ETGs.  Based
on high resolution images obtained by {\it HST} ($Hubble$ $Space$ $Telescope$),
it was found that the central surface brightness profiles of ETGs could be
fitted by a ``Nuker Law'' with the form of $\Sigma(\rm r)\sim {\rm
r}^{-\gamma}$ (Crane et al. 1993; Ferrarese et al. 1994; Lauer et al. 1995).
ETGs with steep inner cusps ($\gamma>0.5$) are classified as ``power-law''
galaxies, while ETGs with shallow inner profiles ($\gamma<0.3$) are called
``core'' galaxies (Lauer et al. 1995; Faber et al. 1997). In a more recent
study, Lauer et al. (2007b) introduced $\gamma'$ as an indicator of the bimodal
classification, which is the local slope at the {\it HST} angular resolution
limit, instead of $\gamma$ that is the inner cusp slope as $r\rightarrow0$.
ETGs with $\gamma'>0.5$ are called ``power-law'' galaxies, ``core''
galaxies are those with $\gamma'<0.3$, and the rest, i.e., ETGs with
$0.3<\gamma'<0.5$ are classified as ``intermediate'' type.
The inner slope of the central profile of ETGs correlates with their global
physical properties such as the luminosity, rotation velocity and
isophotal shape (Faber et al. 1997; Lauer et al. 2007b).

The kinematics of ETGs were usually described by a ratio of the rotational
velocity to the velocity dispersion ($v/\sigma$). Recently, the SAURON Survey
proposed a new tracer of the kinematical properties of ETGs, the specific
angular momentum $\lambda_R$ (see section 3) to divide ETGs into fast and
slow rotators (e.g. de Zeeuw et al. 2002; Emsellem et al. 2007).  As an
extension of the SAURON survey, the \atlas\ team conducted a multi-wavelength
survey for a carefully selected volume-limited ETGs sample with 260 objects
using the SAURON integral-field spectrograph (IFS). Based on these
observations, the \atlas\ team quantitatively classified ETGs into fast and
slow rotators by using $\lambda_R=0.1$ (e.g. Krajnovi\' c et al. 2011).
Fast rotators have regular stellar rotation with alignments between the
photometric and kinematic axes, low luminosity and large ellipticity,
while slow rotators show little or no rotation, and tend to be more massive
and rather round (e.g. Cappellari et al. 2007; Emsellem et al. 2007, 2011).

Given that dichotomies have been found in isophotal shapes, nuclear light
profiles and kinematics, it is interesting to investigate relations between
these properties and with other galactic properties. Virtually, several
studies have focused on such issues but reached conflicting conclusions.
Krajnovi\' c et al. (2013) compared the
nuclear light profiles and large scale kinematics of 135 ETGs, and concluded
that there is no evidence for bimodal distribution in the nuclear slope.
Emsellem et al. (2011) also pointed out that the $a_{4}/a$ parameter appears
not to be directly related to the kinematic properties of ETGs. In a word,
\atlas\ team argues against the dichotomy of ETGs based on their isophotal
shape ($a_{4}/a$) and nuclear light profile.  However, Lauer (2012)
investigated the relation between the kinematics and central structures based
on an ETG sample with 63 objects and found that they are well correlated if a
criterion of $\lambda_{R_e/2}=0.25$ (see section 3) is used to separate fast
and slow rotators. The correlation shows that slow rotator ETGs usually
have cores, while ``power-law'' galaxies tend to rotate rapidly.

From a theoretical point-of-view, simulations of galaxy formation indicate that
the bimodality of isophotal shapes and central profiles of ETGs is correlated
with galaxy merger histories, but such relations are complicated. The
dissipationless simulations by Naab \& Burkert (2003) and Naab \& Trujillo
(2006) showed that the equal-mass mergers of two disk galaxies tend to produce
boxy ETGs, while unequal-mass mergers lead to disky ETGs. But Khochfar \&
Burkert (2005) found that the isophotal shapes of merger remnants depend not
only on the mass ratio of the last major merger, but also on the morphology of
their progenitors and the subsequent gas infall. Using hydrodynamical
simulations, Hopkins et al. (2009a, b) concluded that ``power-law'' ETGs
are formed by dissipational mergers (wet-mergers) in the sense that the inner
extra light/outer profile are formed in a compact central starburst/outer
violent relaxation respectively, whereas ``core'' galaxies are formed by
dry-mergers through subsequent merging of gas-poor ellipticals. During the
process of dry-merging, the center becomes dense and compact because the
merging binary black holes scatter out the inner stars.

Therefore, there is still a debate on the dichotomy of ETGs and their formation history from
both observational and theoretical sides. In this work, we re-investigate the correlations among isophotal
shapes, central light profiles and kinematic properties of ETGs based
on a large compiled sample with 184 ETGs observed by both {\it HST} and SDSS DR8.

The paper is structured as follows. In section 2 we describe the 
sample used for this work. Then we outline the data reduction in section 3.
We present the main results in section 4 and finish with a summary in section 5.
We adopt a Hubble constant of ${\rm H}_{\rm 0}=70\,{\rm km \, s^{-1} Mpc^{-1}}$,
a cosmology with matter density parameter $\Omega_{m}=0.3$,
and a cosmological constant $\Omega_{\Lambda}=0.7$.

\section{Sample}
\label{sect:sample}

To explore the relations between isophotal shapes, central light profiles and
kinematics of ETGs, we need an ETG sample with these properties either
available or measurable.  As mentioned in the introduction, in the literature
there are large ETGs samples observed with {\it HST} and their central light
profiles have been carefully investigated. However, these observations only
cover the central parts of the ETGs because of the small field of view of {\it
HST}. Therefore SDSS images will be used to measure the global properties of
ETGs instead.

Our ETGs sample is compiled from three sources. The first is from the
cross-correlation of the SDSS DR8 photometric catalog with 219 ETGs collected by
Lauer et al. (2007b).  The sample galaxies in Lauer et al. (2007b) were
observed by {\it HST}, WFPC2 ($36\arcsec.5 \times 36\arcsec.5$,
$0\arcsec.046$/pixel, Rest et al. 2001; Laine et al. 2003; Lauer et al. 2005),
WFPC1 ($66\arcsec \times 66\arcsec$, $0\arcsec.043$/pixel, Lauer et al. 1995;
Faber et al. 1997) and NICMOS ($19\arcsec.2 \times 19\arcsec.2$,
$0\arcsec.076$/pixel, Quillen et al. 2000; Ravindranath et al. 2001).  It
encompasses 117 ``core'' galaxies, 89 ``power-law'' galaxies and 13
``intermediate'' galaxies.  The cross-correlation of these 219 ETGs with the
SDSS DR8 leads to 111 ETGs, in which there are 54 ``core'' galaxies, 54
``power-law'' galaxies and 3 ``intermediate'' galaxies, respectively.
The second is taken from Krajnovi\' c et al. (2013) with 135 \atlas\ galaxies
available in the {\it HST} archive. But 61 out of 135 ETGs have been included in
Lauer et al. (2007b). A cross-correlation of the remaining 74 objects with the
SDSS DR8 photometric catalog leaves us with 52 ETGs, consisting of 3 ``core'' galaxies,
37 ''power-law'' galaxies, and 12 ''intermediate'' galaxies.
The third is from the cross-correlation of the SDSS DR8 photometric catalogue with
the sample of 23 ETGs studied by Hyde et al. (2008), which were observed with
Advanced Camera for Surveys (ACS) on {\it HST} and the velocity dispersions are
larger than $350$ {\it km/s}.  21 ETGs were selected, including 6 ``core''
galaxies, 9 ``power-law'' galaxies and 6 ``intermediate'' galaxies.  In total,
we construct an ETGs sample with 184 galaxies, which consist of 63 ``core''
galaxies, 100 ``power-law'' galaxies and 21 ``intermediate'' galaxies,
respectively. The redshifts of 111 ETGs from Lauer et al. (2007b) and 52 ETGs from Krajnovi\' c et al. (2013)
are less than 0.04, while the 21 ETGs from Hyde et al. (2008) are in the range of
$0.1<z<0.3$.  All ETGs are in the luminosity range of $-24<M_V<-15$.

\section{Data Reduction And Parameter Estimation}
\label{sect:data}

We obtained the corrected frame fpC-images in the $r$-band for our sample ETGs
directly from the SDSS DR8 Data Archive Server. For each frame
($2048 \times 1489$ pixels), reductions including bias subtraction,
flat-fielding, pixels defects and cosmic rays correction have been performed
by the SDSS photometric pipeline (PHOTO, Lupton et al. 2001).

The background subtraction process is similar to that of Liu et al. (2008)
and He et al. (2013), which has been successfully applied to the brightest ETGs.
In the following, we outline this approach briefly. First, SExtractor
(Bertin \& Amounts 1996) has been used to generate a background-only image with
all detected objects flagged out. Then a median filter with $51 \times 51$
pixels is used to convolve the background-only image. After the median filtering,
second-order Legendre polynomials are fitted to rows and columns separately by
using the IRAF/NFIT1D task. Finally, we obtain the sky background model by using
a circular Gaussian filter with $\sigma=9$ pixels to smooth the fitted frame.
This sky background model is then subtracted from the original SDSS corrected frame.
After sky background subtraction, the frame is trimmed to $501 \times 501$ pixels
with the target galaxy centered and other objects masked out by using SExtractor.
In the following, isophotal photometry will be performed on this final trimmed frame.

The IRAF/ELLIPSE task is used to perform the surface photometry. Given some initial
guesses for galaxy geometric centre, ellipticity, semi-major axis length and 
position angle, the task fits the isophotes by a series of elliptical annuli from
the centre to the outskirts, with a logarithmic step of 0.1 along the semi-major 
axis. The output of IRAF/ELLIPSE includes the mean isophotal intensity, the position
angle P.A., the ellipticity $\epsilon$ for each annulus, and particularly the fourth
harmonic deviations of Fourier analyses from the isophotal ellipses as a function of
the semi-major axis. We derive the characteristic parameters $a_{4}/a$ and 
ellipticity $\epsilon$ by weighting them with the flux within the elliptical annulus 
over a region of twice FWHMs (full width of half-maximum) of seeing to the effective radius $R_e$.

Apart from $a_{4}/a$ and $\epsilon$, several other galactic properties including the
S\' ersic index, luminosity and dynamical mass are calculated.
The S\' ersic index $n$ and effective radius $R_e$ were obtained by fitting a 
point-spread function (PSF) convolved S\' ersic model (S\' ersic 1968) to the $r$-band sky-subtracted
images using the algorithm GALFIT (Peng et al. 2002). 
The absolute magnitude is derived by $M=m-5\rm log(D_{\rm L}/10\,\rm pc)-A-k$,
where the apparent Petrosian magnitude $m$ and the extinction $A$ are taken from
the SDSS DR8 photometric catalogue, $D_{\rm L}$ is the luminosity distance and
the k-correction $k$ is derived using the KCORRECT algorithm of Blanton \& Roweis (2007).
In order to transform the SDSS photometric data to the standard UBVRI Vega magnitude
system, formula of Smith et al. (2002) has been used to get the absolute galaxy
luminosities in the $V$-band.
We derive the dynamical mass of sample ETGs based on the formula of
$M_{dyn}\approx\sigma^{2}R_{e}/G$, where $\sigma$ is the corrected velocity
dispersion at effective radius $R_{e}$ following von der Linden et al. (2007).
In our sample ETGs, the velocity dispersions $\sigma$ are available for 136 objects,
including the center velocity dispersions for 101 ETGs obtained from Lauer et al. (2007a) and the velocity 
dispersions of other 35 ETGs from the SDSS DR8 spectroscopic catalogue.

To compare with the central surface brightness profile and kinematic properties, 
the characteristic parameters $\gamma'$, $r_{\gamma}$ and $\lambda_{Re/2}$ were taken from the 
literature (Lauer et al. 2007a, b; Hyde et al. 2008;  Cappellari et al. 2011; Krajnovi\' c et al. 2013). 
We describe briefly the way they were derived here.
The central surface brightness profile is fitted by a ``Nuker Law'' with the following form 
\begin{equation}
\label{e:Nuker}
I (r)  = 2^{(\beta - \gamma)/\alpha}  I_b \left(\frac{r_b}{r}\right)^\gamma \left[ 1 + \left(\frac{r}{r_b}\right)^{\alpha}\right]^{(\gamma- \beta)/\alpha}\\ ,
\end{equation}
where the break radius $r_b$ is the point of maximum curvature in log-log 
coordinates, $I_b$ is the surface brightness at $r_b$, $-\beta$ is the 
asymptotic outer slop, $\alpha$ is the sharpness of the break, and 
$\gamma$ is the inner cusp slope as $r \rightarrow 0$ and is distinguished from 
$\gamma^\prime$, which is the local slope evaluated at the HST angular resolution limit $r_0$, 
where  
\begin{equation}
\label{e:gamma}
\gamma^\prime \equiv - \frac{d~log~I}{d~log~r} \bigg|_{r = r_0}   = - \frac{\gamma + \beta(r_0/r_b)^\alpha}
{1 + (r_0/r_b)^\alpha} \\ .
\end{equation}
As described in the introduction,
the ETGs are classified into disky ($a_{4}/a>0$) and boxy ($a_{4}/a<0$) galaxies by
their isophotal shapes, and they are also divided into ``core'' ($\gamma'<0.3$),
``power-law'' ($\gamma'>0.5$) and ``intermediate'' ($0.3<\gamma'<0.5$) galaxies
according to their central light profiles.
For ``core'' galaxies, the physical scale of the core is characterized by the ``cusp radius'', 
$r_{\gamma}$, which is a radius at which $\gamma'$ equals 0.5. Specifically, $r_{\gamma}$ is given by
\begin{equation}
\label{e:cusp}
r_\gamma \equiv r_b \left(  \frac{0.5 - \gamma}{\beta - 0.5}\right)^{1/\alpha}\\ .
\end{equation}
The specific angular momentum parameter $\lambda_R$ is used as a discriminator of 
fast and slow rotators, where $\lambda_R$ is defined as
\begin{equation}
\label{eq:sumLambda}
\lambda_R = \frac{\sum_{n=1}^{N} F_n R_n \left| V_n \right|}{\sum_{n=1}^{N} F_n R_n \sqrt{V_n^2+\sigma_n^2}} \, ,
\end{equation}
where $F_n$ is the flux, $R_n$ is the circular radius from the center of the galaxy,
$V_n$ and $\sigma_n$ are velocity and velocity dispersion inside the $n$-th spatial radial bin.
Particularly, $\lambda_{R_e/2}$ is the $\lambda_R$ measured within half of the effective radius $R_e$.

We list all these parameters for ``core'', ``power-law'' and ``intermediate'' galaxies
in Tables~\ref{Table1}, ~\ref{Table2} and ~\ref{Table3}, respectively.

%%%%%%%%%%%%%%%%%%%%%%%%%%%%%tables
\begin{center}
\small
\setlength\LTleft{0pt}
\setlength\LTright{0pt}
%\begin{landscape}
\begin{longtable}{cccccccccccc}
\caption{``Core'' Galaxy Parameters\label{Table1}}\\
  \hline\noalign{\smallskip}
N& Galaxy& $\gamma'$& $r_\gamma$& $R_{e}$& $\sigma$& $a_{4}/a$& $\varepsilon$& $M_V$ & $\log {M_{dyn}\over M_{\odot}}$ & $n$ & $\lambda_{R_e/2}$ \\
 &  & & $\log({\rm pc})$ & $\log({\rm pc})$ & ${\rm km \, \rm s^{-1}}$ & $10^{(-2)}$ & & ${\rm mag}$ & & & \\
(1) & (2) & (3) & (4) & (5) & (6) & (7) & (8) & (9) & (10) &(11) & (12) \\

  \hline\noalign{\smallskip}

1 & IC 0613 & 0.25$^b$ & 2.05$^a$ & 3.97 & 262$^a$ & 0.125$\pm$0.010 & 0.084$\pm$0.002 & -22.27 & 11.17 & 5.26 & -- \\
2 & IC 0664 & 0.12$^b$ & 2.07$^a$ & 4.84 & 336$^a$ & -0.286$\pm$0.014 & 0.229$\pm$0.002 & -22.86 & 12.26 & 8.87 & -- \\
3 & IC 0712 & 0.17$^b$ & 2.69$^a$ & 4.75 & 345$^a$ & 0.132$\pm$0.006 & 0.188$\pm$0.001 & -23.29 & 12.19 & 7.79 & -- \\
4 & IC 1565 & -0.03$^b$ & 1.65$^a$ & 3.75 & 303$^a$ & -0.008$\pm$0.074 & 0.046$\pm$0.002 & -22.99 & 11.08 & 14.59 & -- \\
5 & IC 1695 & 0.23$^b$ & 2.36$^a$ & 4.74 & 364$^a$ & 0.049$\pm$0.097 & 0.234$\pm$0.004 & -23.90 & 12.23 & 8.22 & -- \\
6 & IC 1733 & -0.01$^b$ & 2.68$^a$ & 4.59 & 301$^a$ & 0.433$\pm$0.088 & 0.126$\pm$0.002 & -23.43 & 11.91 & 4.87 & -- \\
7 & J010803.2+151333.6 & 0.23$^d$ & -- & 4.22 & 304$^e$ & -0.026$\pm$0.054 & 0.170$\pm$0.007 & -23.19 & 11.55 & 3.28 & -- \\
8 & J083445.2+355142.0 & 0.06$^d$ & -- & 4.60 & 366$^e$ & 0.067$\pm$0.088 & 0.175$\pm$0.007 & -23.71 & 12.09 & 4.96 & -- \\
9 & J124609.4+515021.6 & 0.21$^d$ & -- & 4.38 & 387$^e$ & 0.509$\pm$0.090 & 0.097$\pm$0.012 & -23.85 & 11.92 & 3.99 & -- \\
10 & J141341.4+033104.3 & -0.09$^d$ & -- & 4.77 & 364$^e$ & -0.119$\pm$0.112 & 0.162$\pm$0.012 & -23.42 & 12.26 & 1.89 & -- \\
11 & J171328.4+274336.6 & 0.04$^d$ & -- & 4.49 & 414$^e$ & 0.865$\pm$0.093 & 0.171$\pm$0.013 & -24.17 & 12.09 & 5.61 & -- \\
12 & J211019.2+095047.1 & 0.17$^d$ & -- & 4.25 & 371$^e$ & -0.981$\pm$0.053 & 0.144$\pm$0.006 & -23.78 & 11.76 & 2.00 & -- \\
13 & MCG 11-14-25A & 0.30$^b$ & 1.38$^a$ & 3.34 & 148$^e$ & 0.185$\pm$0.007 & 0.097$\pm$0.001 & -19.08 & 10.05 & 4.34 & -- \\
14 & NGC 0524 & 0.27$^b$ & 1.57$^a$ & 3.50 & 253$^a$ & 0.218$\pm$0.035 & 0.034$\pm$0.001 & -21.85 & 10.67 & 4.55 & 0.325 \\
15 & NGC 0545 & 0.10$^b$ & 2.16$^a$ & 4.36 & 242$^a$ & 0.462$\pm$0.059 & 0.239$\pm$0.001 & -22.98 & 11.49 & 16.99 & -- \\
16 & NGC 0584 & 0.30$^b$ & 0.95$^a$ & 3.53 & 207$^a$ & 0.106$\pm$0.028 & 0.250$\pm$0.001 & -21.38 & 10.53 & 7.06 & -- \\
17 & NGC 0741 & 0.11$^b$ & 2.46$^a$ & 4.12 & 291$^a$ & 0.090$\pm$0.037 & 0.128$\pm$0.001 & -23.27 & 11.41 & 6.18 & -- \\
18 & NGC 1016 & 0.11$^b$ & 2.25$^a$ & 2.10 & 294$^a$ & -0.026$\pm$0.035 & 0.066$\pm$0.001 & -22.90 & 9.40 & 7.74 & -- \\
19 & NGC 1052 & 0.22$^b$ & 1.46$^a$ & 3.45 & 208$^a$ & -0.773$\pm$0.001 & 0.265$\pm$0.001 & -21.17 & 10.45 & 2.75 & -- \\
20 & NGC 1700 & 0.07$^b$ & 1.01$^a$ & 3.64 & 235$^a$ & 0.986$\pm$0.029 & 0.266$\pm$0.001 & -21.95 & 10.75 & 12.12 & -- \\
21 & NGC 2832 & 0.03$^b$ & 2.52$^a$ & 4.93 & 335$^a$ & -0.333$\pm$0.003 & 0.192$\pm$0.001 & -23.76 & 12.35 & 9.08 & -- \\
22 & NGC 3193 & 0.28$^b$ & 1.38$^a$ & 3.49 & 194$^a$ & 0.317$\pm$0.001 & 0.161$\pm$0.001 & -21.98 & 10.43 & 4.92 & 0.197 \\
23 & NGC 3379 & 0.18$^b$ & 1.72$^a$ & 3.67 & 207$^a$ & -0.028$\pm$0.001 & 0.098$\pm$0.001 & -21.14 & 10.67 & 6.66 & 0.157 \\
24 & NGC 3551 & 0.14$^b$ & 2.37$^a$ & 5.44 & 268$^a$ & 0.374$\pm$0.005 & 0.173$\pm$0.001 & -23.55 & 12.66 & 9.43 & -- \\
25 & NGC 3607 & 0.26$^b$ & 1.77$^a$ & 3.51 & 224$^a$ & -0.099$\pm$0.001 & 0.192$\pm$0.001 & -19.88 & 10.58 & 4.61 & 0.228 \\
26 & NGC 3608 & 0.17$^b$ & 1.31$^a$ & 3.73 & 193$^a$ & -0.420$\pm$0.001 & 0.175$\pm$0.001 & -21.12 & 10.67 & 5.71 & 0.043 \\
27 & NGC 3613 & 0.08$^b$ & 1.65$^a$ & 3.56 & 210$^a$ & -0.123$\pm$0.003 & 0.313$\pm$0.001 & -21.59 & 10.57 & 2.92 & 0.191 \\
28 & NGC 3640 & 0.03$^b$ & 1.47$^a$ & 3.41 & 182$^a$ & -0.305$\pm$0.001 & 0.214$\pm$0.001 & -21.96 & 10.30 & 3.41 & 0.320 \\
29 & NGC 3842 & 0.12$^b$ & 2.48$^a$ & 4.41 & 314$^a$ & -0.387$\pm$0.003 & 0.149$\pm$0.001 & -23.18 & 11.77 & 5.59 & -- \\
30 & NGC 4073 & -0.08$^b$ & 2.13$^a$ & 4.56 & 278$^a$ & 0.349$\pm$0.003 & 0.297$\pm$0.001 & -23.50 & 11.81 & 5.16 & -- \\
31 & NGC 4168 & 0.17$^b$ & 2.26$^a$ & 3.76 & 184$^a$ & 0.804$\pm$0.002 & 0.155$\pm$0.001 & -21.80 & 10.66 & 3.61 & 0.040 \\
32 & NGC 4261 & 0.00$^b$ & 2.31$^a$ & 3.96 & 309$^a$ & -1.372$\pm$0.001 & 0.256$\pm$0.001 & -22.26 & 11.31 & 5.31 & 0.085 \\
33 & NGC 4278 & 0.10$^b$ & 1.77$^a$ & 3.16 & 238$^a$ & -0.280$\pm$0.001 & 0.148$\pm$0.001 & -21.05 & 10.28 & 4.49 & 0.203 \\
34 & NGC 4365 & 0.09$^b$ & 2.15$^a$ & 4.06 & 256$^a$ & -1.181$\pm$0.001 & 0.238$\pm$0.001 & -22.18 & 11.24 & 6.26 & 0.088 \\
35 & NGC 4371 & 0.27$^c$ & 1.60$^c$ & 3.24 & -- & 0.512$\pm$0.061 & 0.257$\pm$0.002 & -20.00 & -- & 3.43 & 0.482 \\
36 & NGC 4374 & 0.13$^b$ & 2.11$^a$ & 3.79 & 282$^a$ & -0.401$\pm$0.001 & 0.183$\pm$0.002 & -22.28 & 11.06 & 5.62 & 0.024 \\
37 & NGC 4382 & 0.01$^b$ & 1.69$^a$ & 4.04 & 179$^a$ & 0.852$\pm$0.001 & 0.212$\pm$0.001 & -21.96 & 10.91 & 6.00 & 0.163 \\
38 & NGC 4406 & -0.04$^b$ & 1.90$^a$ & 3.34 & 235$^a$ & -0.763$\pm$0.001 & 0.180$\pm$0.001 & -22.46 & 10.45 & 6.68 & 0.052 \\
39 & NGC 4458 & 0.17$^b$ & 0.80$^a$ & 3.51 & 103$^a$ & 0.395$\pm$0.003 & 0.138$\pm$0.001 & -19.27 & 9.90 & 6.97 & 0.079 \\
40 & NGC 4472 & 0.01$^b$ & 2.25$^a$ & 3.56 & 291$^a$ & -0.227$\pm$0.001 & 0.087$\pm$0.001 & -22.93 & 10.85 & 3.01 & 0.077 \\
41 & NGC 4473 & 0.01$^b$ & 1.73$^a$ & 3.67 & 179$^a$ & 1.149$\pm$0.001 & 0.388$\pm$0.001 & -21.16 & 10.54 & 4.28 & 0.250 \\
42 & NGC 4478 & 0.10$^b$ & 1.32$^a$ & 3.08 & 138$^a$ & -0.449$\pm$0.002 & 0.181$\pm$0.001 & -19.89 & 9.73 & 1.84 & 0.177 \\
43 & NGC 4486 & 0.27$^b$ & 2.65$^a$ & 3.53 & 332$^a$ & -0.098$\pm$0.001 & 0.017$\pm$0.001 & -22.71 & 10.94 & 2.14 & -- \\
44 & NGC 4486B & -0.10$^b$ & 1.08$^a$ & 2.49 & 170$^a$ & 0.458$\pm$0.005 & 0.110$\pm$0.002 & -17.98 & 9.32 & 2.10 & 0.021 \\
45 & NGC 4552 & -0.02$^b$ & 1.60$^a$ & 2.89 & 253$^a$ & -0.010$\pm$0.001 & 0.050$\pm$0.001 & -21.65 & 10.06 & 4.43 & 0.049 \\
46 & NGC 4636 & 0.13$^b$ & 2.21$^a$ & 3.59 & 203$^a$ & -0.018$\pm$0.001 & 0.026$\pm$0.001 & -21.86 & 10.57 & 3.44 & 0.036 \\
47 & NGC 4649 & 0.17$^b$ & 2.34$^a$ & 3.62 & 336$^a$ & -0.477$\pm$0.001 & 0.113$\pm$0.001 & -22.51 & 11.04 & 3.23 & 0.127 \\
48 & NGC 4874 & 0.12$^b$ & 2.99$^a$ & 4.80 & 278$^a$ & -0.058$\pm$0.003 & 0.074$\pm$0.001 & -23.49 & 12.05 & 4.96 & -- \\
49 & NGC 4889 & 0.03$^b$ & 2.84$^a$ & 4.11 & 401$^a$ & -0.563$\pm$0.002 & 0.268$\pm$0.001 & -23.73 & 11.68 & 3.41 & -- \\
50 & NGC 5198 & 0.26$^b$ & 1.33$^a$ & 3.60 & 196$^a$ & -0.249$\pm$0.003 & 0.130$\pm$0.001 & -21.23 & 10.55 & 3.53 & 0.057 \\
51 & NGC 5322 & 0.15$^c$ & 2.02$^c$ & 3.64 & -- & -0.001$\pm$0.029 & 0.320$\pm$0.001 & -21.41 & -- & 6.11 & 0.067 \\
52 & NGC 5485 & 0.19$^c$ & 1.90$^c$ & 3.75 & 176$^e$ & -0.599$\pm$0.450 & 0.068$\pm$0.010 & -21.14 & 10.61 & 5.09 & 0.149 \\
53 & NGC 5557 & 0.07$^b$ & 1.82$^a$ & 3.99 & 254$^a$ & -0.274$\pm$0.002 & 0.202$\pm$0.001 & -22.62 & 11.17 & 5.33 & 0.045 \\
54 & NGC 5576 & 0.26$^b$ & 1.21$^a$ & 3.36 & 183$^a$ & -0.642$\pm$0.001 & 0.258$\pm$0.001 & -21.31 & 10.25 & 4.65 & 0.091 \\
55 & NGC 5813 & 0.06$^b$ & 1.89$^a$ & 4.72 & 239$^a$ & 0.042$\pm$0.001 & 0.095$\pm$0.001 & -22.01 & 11.84 & 8.50 & 0.071 \\
56 & NGC 5982 & 0.05$^b$ & 1.80$^a$ & 3.81 & 240$^a$ & -1.241$\pm$0.002 & 0.281$\pm$0.001 & -21.97 & 10.94 & 4.92 & -- \\
57 & NGC 6086 & 0.02$^b$ & 2.53$^a$ & 4.57 & 336$^a$ & -0.562$\pm$0.005 & 0.268$\pm$0.001 & -23.11 & 11.99 & 7.36 & -- \\
58 & NGC 6166 & 0.12$^b$ & 3.17$^a$ & 4.40 & 310$^a$ & -0.372$\pm$0.005 & 0.203$\pm$0.002 & -23.80 & 11.75 & 2.75 & -- \\
59 & NGC 6173 & 0.02$^b$ & 2.32$^a$ & 4.60 & 278$^a$ & -0.334$\pm$0.004 & 0.332$\pm$0.001 & -23.59 & 11.85 & 6.87 & -- \\
60 & NGC 7578B & 0.21$^b$ & 2.06$^a$ & 4.57 & 214$^a$ & 0.499$\pm$0.085 & 0.170$\pm$0.002 & -23.41 & 11.60 & 16.47 & -- \\
61 & NGC 7619 & 0.01$^b$ & 2.03$^a$ & 3.97 & 322$^a$ & 0.238$\pm$0.036 & 0.231$\pm$0.001 & -22.94 & 11.35 & 6.25 & -- \\
62 & NGC 7647 & 0.05$^b$ & 2.28$^a$ & 3.00 & 282$^a$ & 0.994$\pm$0.294 & 0.290$\pm$0.035 & -23.97 & 10.27 & 15.58 & -- \\
63 & NGC 7785 & 0.06$^b$ & 1.32$^a$ & 3.62 & 245$^a$ & -1.708$\pm$0.040 & 0.388$\pm$0.001 & -22.08 & 10.76 & 4.84 & -- \\

\noalign{\smallskip}\hline
\end{longtable}
%\tablenotes{a}{\textwidth}{From Lauer et al. (2007a).}
%\tablenotes{b}{\textwidth}{From Lauer et al. (2007b).}
%\tablenotes{c}{\textwidth}{From Krajnovi\'c et al. (2013).}
%\tablenotes{d}{\textwidth}{From Hyde et al. (2008).}
%\tablenotes{e}{\textwidth}{From the SDSS DR8 Spectroscopic catalogue.}
\tablecomments{\textwidth}{Col.(1): Number. Col.(2): Galaxy Name.
Cols.(3) and (4): The local slope of ``Nuker Law'' fits and ``Cusp radius'',
$^a$ from Lauer et al. (2007a); $^b$ from Lauer et al. (2007b); 
$^c$ from Krajnovi\'c et al. (2013); $^d$ from Hyde et al. (2008).
Col.(5): Effective radius from best S\' ersic fits.
Col.(6): Central velocity dispersion,
$^d$ from Hyde et al. (2008);
$^e$ from the SDSS DR8 Spectroscopic catalogue.
Col.(7): Isophotal shape parameter $a_4/a$.
Col.(8): Ellipticity.
Col.(9): Absolute magnitude in the $V$-band.
Col.(10): The dynamical mass.
Col.(11): S\' ersic index.
Col.(12): Specific angular momentum parameter from Emsellem et al. (2011).
}

%%%%%%%%%%%%%%%%%%%%%%%%%%%%%%%%%%%%%%%%%%%%%%%
\small
\setlength\LTleft{0pt}
\setlength\LTright{0pt}
\begin{longtable}{cccccccccccc}
\caption{``Power-law'' Galaxy Parameters\label{Table2}}\\
 \hline\noalign{\smallskip}
N& Galaxy& $\gamma'$& $r_\gamma$& $R_{e}$& $\sigma$& $a_{4}/a$& $\varepsilon$& $M_V$ & $\log {M_{dyn}\over M_{\odot}}$  & $n$ & $\lambda_{R_e/2}$ \\
 &  & & $\log({\rm pc})$ & $\log({\rm pc})$ & ${\rm km \, \rm s^{-1}}$ & $10^{(-2)}$ & & ${\rm mag}$ & & & \\
(1) & (2) & (3) & (4) & (5) & (6) & (7) & (8) & (9) & (10) &(11) & (12) \\
\hline\noalign{\smallskip}

1 & IC 0875 & 1.12$^b$ & 1.01$^a$ & 4.27 & -- & 0.679$\pm$0.006 & 0.394$\pm$0.001 & -20.21 & -- & 11.20 & -- \\
2 & IC 2738 & 0.60$^b$ & 1.57$^a$ & 4.04 & 275$^a$ & 0.202$\pm$0.009 & 0.068$\pm$0.002 & -22.18 & 11.29 & 5.84 & -- \\
3 & J013431.5+131436.4 & 0.54$^d$ & -- & 4.17 & 248$^e$ & -0.214$\pm$0.309 & 0.524$\pm$0.024 & -23.17 & 11.33 & 9.65 & -- \\
4 & J082216.5+481519.1 & 0.94$^d$ & -- & 5.90 & 351$^e$ & -0.170$\pm$0.155 & 0.305$\pm$0.012 & -21.26 & 13.36 & 15.00 & -- \\
5 & J082646.7+495211.5 & 1.14$^d$ & -- & 4.47 & -- & 0.113$\pm$0.160 & 0.305$\pm$0.014 & -22.10 & -- & 4.83 & -- \\
6 & J093124.4+574926.6 & 0.52$^d$ & -- & 4.17 & 350$^e$ & 0.988$\pm$0.117 & 0.235$\pm$0.012 & -22.95 & 11.62 & 2.20 & -- \\
7 & J103344.2+043143.5 & 0.80$^d$ & -- & 4.12 & 335$^e$ & -0.258$\pm$0.117 & 0.414$\pm$0.010 & -22.36 & 11.54 & 3.36 & -- \\
8 & J111525.7+024033.9 & 0.76$^d$ & -- & 4.34 & 379$^e$ & -0.833$\pm$0.108 & 0.265$\pm$0.011 & -23.40 & 11.86 & 3.21 & -- \\
9 & J151741.7-004217.6 & 1.10$^d$ & -- & 3.93 & 380$^e$ & 0.115$\pm$0.059 & 0.215$\pm$0.005 & -21.67 & 11.46 & 3.10 & -- \\
10 & J160239.1+022110.0 & 0.61$^d$ & -- & 4.00 & 358$^e$ & 1.654$\pm$0.198 & 0.271$\pm$0.017 & -22.91 & 11.47 & 3.08 & -- \\
11 & J221414.3+131703.7 & 1.09$^d$ & -- & 3.60 & -- & 1.689$\pm$0.094 & 0.274$\pm$0.009 & -21.85 & -- & 2.18 & -- \\
12 & MCG 08-27-18 & 0.89$^b$ & 1.07$^a$ & 3.18 & 89$^a$ & -0.074$\pm$0.006 & 0.076$\pm$0.001 & -20.03 & 9.45 & 3.41 & -- \\
13 & NGC 0474 & 0.56$^b$ & 1.15$^a$ & 3.59 & 164$^a$ & -0.212$\pm$0.043 & 0.122$\pm$0.001 & -20.12 & 10.39 & 9.24 & 0.210 \\
14 & NGC 0596 & 0.54$^b$ & 0.63$^a$ & 3.51 & 152$^a$ & 0.034$\pm$0.029 & 0.069$\pm$0.001 & -20.90 & 10.24 & 8.28 & -- \\
15 & NGC 0936 & 0.52$^c$ & 0.87$^c$ & 3.50 & -- & 0.145$\pm$0.025 & 0.108$\pm$0.001 & -20.84 & -- & 6.13 & 0.430 \\
16 & NGC 2549 & 0.67$^b$ & 0.51$^a$ & 3.25 & 143$^a$ & 1.938$\pm$0.003 & 0.447$\pm$0.001 & -19.17 & 9.93 & 3.18 & 0.523 \\
17 & NGC 2592 & 0.92$^b$ & 0.82$^a$ & 3.24 & 265$^a$ & 0.579$\pm$0.003 & 0.158$\pm$0.001 & -20.01 & 10.45 & 3.53 & 0.431 \\
18 & NGC 2685 & 0.73$^b$ & 0.84$^a$ & 3.23 & 94$^a$ & 3.180$\pm$0.004 & 0.533$\pm$0.002 & -19.72 & 9.54 & 3.59 & 0.632 \\
19 & NGC 2778 & 0.83$^b$ & 0.67$^a$ & 3.31 & 162$^a$ & 0.721$\pm$0.004 & 0.208$\pm$0.001 & -18.75 & 10.10 & 1.86 & 0.435 \\
20 & NGC 2859 & 0.76$^c$ & 0.77$^c$ & 3.18 & -- & 1.109$\pm$0.035 & 0.188$\pm$0.001 & -20.83 & -- & 3.60 & 0.361 \\
21 & NGC 2872 & 1.01$^b$ & 1.06$^a$ & 3.65 & 285$^a$ & -0.147$\pm$0.002 & 0.200$\pm$0.001 & -21.62 & 10.93 & 4.15 & -- \\
22 & NGC 2880 & 0.75$^c$ & 1.01$^c$ & 3.29 & 281$^e$ & -0.215$\pm$0.035 & 0.210$\pm$0.001 & -20.31 & 10.56 & 4.95 & 0.482 \\
23 & NGC 2950 & 0.82$^b$ & 0.58$^a$ & 3.27 & 182$^a$ & 0.819$\pm$0.002 & 0.242$\pm$0.001 & -19.73 & 10.16 & 5.17 & 0.428 \\
24 & NGC 2962 & 0.80$^c$ & 1.21$^c$ & 3.77 & -- & 1.509$\pm$0.076 & 0.280$\pm$0.001 & -20.42 & -- & 8.37 & 0.329 \\
25 & NGC 3156 & 1.78$^c$ & 1.02$^c$ & 3.85 & -- & -0.252$\pm$0.062 & 0.435$\pm$0.001 & -19.36 & -- & 8.62 & 0.559 \\
26 & NGC 3226 & 0.83$^c$ & 0.57$^c$ & 3.87 & -- & -0.267$\pm$0.048 & 0.162$\pm$0.001 & -19.59 & -- & 9.59 & 0.257 \\
27 & NGC 3245 & 0.74$^c$ & 0.99$^c$ & 3.63 & 850$^e$ & 0.476$\pm$0.032 & 0.318$\pm$0.001 & -20.62 & 11.86 & 8.60 & 0.592 \\
28 & NGC 3266 & 0.66$^b$ & 0.85$^a$ & 3.40 & -- & 1.889$\pm$0.005 & 0.117$\pm$0.002 & -20.11 & -- & 7.15 & -- \\
29 & NGC 3377 & 0.62$^b$ & 0.36$^a$ & 3.37 & 139$^a$ & 0.446$\pm$0.001 & 0.343$\pm$0.006 & -20.07 & 10.02 & 4.55 & 0.522 \\
30 & NGC 3384 & 0.71$^b$ & 0.36$^a$ & 4.22 & 148$^a$ & 0.995$\pm$0.001 & 0.251$\pm$0.001 & -19.93 & 10.93 & 14.90 & 0.397 \\
31 & NGC 3412 & 0.67$^c$ & 0.73$^c$ & 3.40 & -- & 0.146$\pm$0.035 & 0.254$\pm$0.001 & -19.96 & -- & 11.29 & 0.403 \\
32 & NGC 3414 & 0.84$^b$ & 0.81$^a$ & 3.82 & 237$^a$ & 1.746$\pm$0.002 & 0.216$\pm$0.001 & -20.25 & 10.94 & 6.07 & 0.070 \\
33 & NGC 3458 & 0.59$^c$ & 1.17$^c$ & 3.18 & -- & -0.146$\pm$0.037 & 0.120$\pm$0.001 & -19.86 & -- & 8.41 & 0.250 \\
34 & NGC 3489 & 0.57$^c$ & 0.72$^c$ & 2.84 & -- & -0.238$\pm$0.043 & 0.250$\pm$0.002 & -19.30 & -- & 4.64 & 0.552 \\
35 & NGC 3595 & 0.76$^b$ & 0.93$^a$ & 3.39 & -- & -0.316$\pm$0.004 & 0.343$\pm$0.001 & -20.96 & -- & 2.87 & 0.301 \\
36 & NGC 3599 & 0.75$^b$ & 0.65$^a$ & 3.77 & 85$^a$ & 0.264$\pm$0.003 & 0.115$\pm$0.001 & -19.93 & 10.00 & 7.45 & 0.239 \\
37 & NGC 3605 & 0.60$^b$ & 0.65$^a$ & 2.52 & 92$^a$ & -0.734$\pm$0.037 & 0.261$\pm$0.001 & -19.61 & 8.81 & 2.23 & 0.347 \\
38 & NGC 3610 & 0.76$^b$ & 0.64$^a$ & 3.28 & 162$^a$ & 2.128$\pm$0.003 & 0.437$\pm$0.002 & -20.96 & 10.07 & 3.96 & 0.539 \\
39 & NGC 3796 & 0.74$^c$ & 1.04$^c$ & 2.88 & -- & 0.125$\pm$0.052 & 0.370$\pm$0.001 & -18.66 & -- & 10.39 & 0.119 \\
40 & NGC 3900 & 1.02$^b$ & 1.16$^a$ & 3.61 & 118$^a$ & 0.294$\pm$0.003 & 0.233$\pm$0.002 & -20.80 & 10.12 & 2.25 & -- \\
41 & NGC 3945 & 0.57$^b$ & 0.59$^a$ & 3.37 & 174$^a$ & 2.645$\pm$0.003 & 0.230$\pm$0.001 & -20.25 & 10.22 & 5.03 & 0.561 \\
42 & NGC 4026 & 0.65$^b$ & 0.48$^a$ & 3.26 & 178$^a$ & 4.249$\pm$0.003 & 0.368$\pm$0.002 & -19.79 & 10.13 & 2.52 & 0.442 \\
43 & NGC 4121 & 0.85$^b$ & 0.79$^a$ & 2.73 & 86$^a$ & -0.085$\pm$0.005 & 0.242$\pm$0.001 & -18.53 & 8.97 & 1.32 & -- \\
44 & NGC 4128 & 0.75$^b$ & 0.92$^a$ & 3.48 & 203$^a$ & -0.498$\pm$0.042 & 0.370$\pm$0.001 & -20.79 & 10.46 & 9.25 & -- \\
45 & NGC 4143 & 0.61$^b$ & 0.88$^a$ & 3.09 & 214$^a$ & 0.714$\pm$0.002 & 0.228$\pm$0.001 & -19.68 & 10.12 & 2.62 & 0.398 \\
46 & NGC 4150 & 0.68$^b$ & 0.85$^a$ & 3.49 & 85$^a$ & 0.084$\pm$0.002 & 0.209$\pm$0.002 & -18.66 & 9.72 & 10.40 & 0.338 \\
47 & NGC 4203 & 0.74$^c$ & 0.85$^c$ & 2.98 & -- & 0.655$\pm$0.028 & 0.082$\pm$0.001 & -19.83 & -- & 4.89 & 0.275 \\
48 & NGC 4262 & 0.76$^c$ & 0.87$^c$ & 2.81 & -- & 0.228$\pm$0.025 & 0.099$\pm$0.001 & -20.04 & -- & 4.31 & 0.250 \\
49 & NGC 4267 & 0.71$^c$ & 0.88$^c$ & 2.92 & -- & 0.769$\pm$0.024 & 0.073$\pm$0.001 & -19.78 & -- & 3.94 & 0.253 \\
50 & NGC 4281 & 0.56$^c$ & 1.04$^c$ & 4.20 & -- & 0.774$\pm$0.041 & 0.463$\pm$0.001 & -21.84 & -- & 12.00 & 0.621 \\
51 & NGC 4283 & 0.80$^c$ & 1.04$^c$ & 2.75 & -- & -0.091$\pm$0.031 & 0.052$\pm$0.001 & -18.72 & -- & 3.39 & 0.151 \\
52 & NGC 4339 & 0.81$^c$ & 0.87$^c$ & 3.59 & -- & -0.020$\pm$0.033 & 0.051$\pm$0.001 & -19.95 & -- & 7.30 & 0.312 \\
53 & NGC 4340 & 0.68$^c$ & 0.89$^c$ & 3.94 & -- & 0.558$\pm$0.131 & 0.192$\pm$0.003 & -19.44 & -- & 12.35 & 0.442 \\
54 & NGC 4342 & 0.55$^c$ & 0.84$^c$ & 3.23 & 219$^e$ & 3.559$\pm$0.102 & 0.445$\pm$0.002 & -18.13 & 10.28 & 8.00 & 0.306 \\
55 & NGC 4387 & 0.65$^b$ & 0.54$^a$ & 2.71 & 104$^a$ & -1.414$\pm$0.003 & 0.304$\pm$0.001 & -19.25 & 9.11 & 2.35 & 0.317 \\
56 & NGC 4417 & 0.75$^b$ & 0.94$^a$ & 3.24 & 131$^a$ & 1.868$\pm$0.002 & 0.345$\pm$0.001 & -18.94 & 9.84 & 4.68 & 0.392 \\
57 & NGC 4429 & 1.07$^c$ & 0.58$^c$ & 4.33 & -- & -0.1344$\pm$0.044 & 0.442$\pm$0.001 & -20.41 & -- & 9.35 & 0.396 \\
58 & NGC 4434 & 0.64$^b$ & 0.54$^a$ & 3.00 & 120$^a$ & 0.124$\pm$0.003 & 0.062$\pm$0.001 & -19.19 & 9.52 & 3.91 & 0.199 \\
59 & NGC 4442 & 0.52$^c$ & 0.72$^c$ & 2.97 & 59$^e$ & -0.622$\pm$0.032 & 0.285$\pm$0.001 & -19.04 & 8.88 & 4.90 & 0.338 \\
60 & NGC 4464 & 0.70$^b$ & 0.54$^a$ & 2.80 & 127$^a$ & 0.714$\pm$0.003 & 0.280$\pm$0.001 & -18.82 & 9.37 & 3.24 & -- \\
61 & NGC 4467 & 0.94$^b$ & 0.54$^a$ & 2.91 & 68$^a$ & 0.606$\pm$0.007 & 0.285$\pm$0.001 & -17.51 & 8.94 & 4.00 & -- \\
62 & NGC 4474 & 0.72$^b$ & 0.72$^a$ & 3.45 & 87$^a$ & 2.271$\pm$0.004 & 0.267$\pm$0.001 & -18.42 & 9.70 & 3.93 & 0.353 \\
63 & NGC 4483 & 0.88$^c$ & 0.91$^c$ & 3.38 & 92$^e$ & -0.245$\pm$0.056 & 0.262$\pm$0.001 & -18.44 & 9.68 & 8.27 & 0.273 \\
64 & NGC 4486A & 0.72$^c$ & 0.95$^c$ & 2.98 & -- & -0.046$\pm$0.072 & 0.244$\pm$0.005 & -18.92 & -- & 5.16 & 0.351 \\
65 & NGC 4489 & 0.64$^c$ & 0.87$^c$ & 4.41 & 62$^e$ & 0.174$\pm$0.061 & 0.088$\pm$0.001 & -18.64 & 10.37 & 9.39 & 0.117 \\
66 & NGC 4494 & 0.55$^b$ & 0.54$^a$ & 3.68 & 150$^a$ & 0.032$\pm$0.001 & 0.148$\pm$0.001 & -21.50 & 10.40 & 3.69 & 0.212 \\
67 & NGC 4503 & 0.65$^b$ & 0.63$^a$ & 4.28 & 111$^a$ & 0.029$\pm$0.002 & 0.270$\pm$0.001 & -19.57 & 10.74 & 7.69 & 0.470 \\
68 & NGC 4528 & 0.97$^c$ & 0.88$^c$ & 2.94 & -- & -0.615$\pm$0.059 & 0.169$\pm$0.002 & -19.72 & -- & 3.72 & 0.102 \\
69 & NGC 4550 & 0.57$^c$ & 0.88$^c$ & 2.59 & -- & 2.086$\pm$0.086 & 0.582$\pm$0.001 & -17.32 & -- & 1.94 & 0.061 \\
70 & NGC 4551 & 0.69$^b$ & 0.54$^a$ & 3.12 & 108$^a$ & -0.456$\pm$0.002 & 0.260$\pm$0.001 & -19.37 & 9.55 & 2.14 & 0.259 \\
71 & NGC 4564 & 0.81$^b$ & 0.63$^a$ & 3.36 & 157$^a$ & 1.426$\pm$0.002 & 0.333$\pm$0.001 & -20.26 & 10.12 & 4.50 & 0.536 \\
72 & NGC 4570 & 0.85$^c$ & 0.92$^c$ & 3.51 & -- & 1.154$\pm$0.073 & 0.396$\pm$0.001 & -21.26 & -- & 6.48 & 0.498 \\
73 & NGC 4578 & 0.89$^c$ & 0.90$^c$ & 3.81 & -- & 0.342$\pm$0.033 & 0.233$\pm$0.001 & -21.01 & -- & 7.66 & 0.544 \\
74 & NGC 4596 & 0.77$^c$ & 0.90$^c$ & 4.04 & -- & 1.289$\pm$0.058 & 0.216$\pm$0.001 & -21.42 & -- & 7.31 & 0.280 \\
75 & NGC 4612 & 0.64$^c$ & 0.91$^c$ & 4.60 & -- & -0.207$\pm$0.041 & 0.192$\pm$0.001 & -20.79 & -- & 16.74 & 0.324 \\
76 & NGC 4621 & 0.85$^b$ & 0.54$^a$ & 3.43 & 225$^a$ & 1.539$\pm$0.001 & 0.325$\pm$0.001 & -21.74 & 10.50 & 6.06 & 0.291 \\
77 & NGC 4623 & 2.06$^c$ & 0.93$^c$ & 3.59 & -- & 1.112$\pm$0.100 & 0.562$\pm$0.001 & -19.80 & -- & 5.02 & 0.564 \\
78 & NGC 4638 & 0.77$^c$ & 0.93$^c$ & 3.04 & -- & 3.801$\pm$0.172 & 0.538$\pm$0.002 & -20.03 & -- & 3.41 & 0.715 \\
79 & NGC 4660 & 0.91$^b$ & 0.54$^a$ & 2.99 & 188$^a$ & 1.100$\pm$0.002 & 0.345$\pm$0.001 & -20.13 & 9.90 & 3.90 & 0.475 \\
80 & NGC 4754 & 0.60$^c$ & 0.04$^c$ & 3.70 & -- & -0.038$\pm$0.032 & 0.182$\pm$0.001 & -20.84 & -- & 8.79 & 0.418 \\
81 & NGC 5173 & 0.52$^c$ & 1.27$^c$ & 3.48 & 99$^e$ & 0.140$\pm$0.036 & 0.121$\pm$0.001 & -20.09 & 9.84 & 9.26 & 0.106 \\
82 & NGC 5273 & 1.66$^c$ & 0.89$^c$ & 3.85 & 90$^e$ & 0.130$\pm$0.052 & 0.135$\pm$0.001 & -19.45 & 10.13 & 10.00 & 0.482 \\
83 & NGC 5308 & 0.96$^b$ & 0.90$^a$ & 3.51 & 211$^a$ & 3.143$\pm$0.004 & 0.467$\pm$0.001 & -21.26 & 10.53 & 3.05 & 0.510 \\
84 & NGC 5370 & 0.67$^b$ & 1.04$^a$ & 3.54 & 133$^a$ & 3.403$\pm$0.010 & 0.294$\pm$0.002 & -20.60 & 10.15 & 4.38 & -- \\
85 & NGC 5831 & 0.55$^b$ & 0.85$^a$ & 3.67 & 164$^a$ & 0.305$\pm$0.002 & 0.272$\pm$0.001 & -21.00 & 10.47 & 5.11 & 0.065 \\
86 & NGC 5838 & 0.93$^b$ & 1.03$^a$ & 3.52 & 266$^a$ & -0.057$\pm$0.002 & 0.167$\pm$0.002 & -20.51 & 10.74 & 4.73 & 0.460 \\
87 & NGC 5845 & 0.52$^b$ & 1.14$^a$ & 2.64 & 234$^a$ & -0.497$\pm$0.002 & 0.255$\pm$0.001 & -19.98 & 9.74 & 2.95 & 0.358 \\
88 & NGC 5854 & 1.01$^c$ & 0.39$^c$ & 3.76 & -- & -0.471$\pm$0.091 & 0.370$\pm$0.001 & -20.41 & -- & 9.08 & 0.515 \\
89 & NGC 6278 & 0.67$^b$ & 0.99$^a$ & 3.66 & 150$^a$ & 0.653$\pm$0.003 & 0.233$\pm$0.001 & -20.81 & 10.38 & 6.26 & 0.411 \\
90 & NGC 6340 & 0.64$^b$ & 0.91$^a$ & 3.56 & 144$^a$ & -0.174$\pm$0.002 & 0.036$\pm$0.001 & -19.46 & 10.24 & 15.97 & -- \\
91 & NGC 7280 & 0.87$^c$ & 1.06$^c$ & 4.17 & -- & -0.485$\pm$0.062 & 0.329$\pm$0.001 & -20.37 & -- & 9.98 & 0.503 \\
92 & NGC 7332 & 0.80$^b$ & 0.67$^a$ & 3.20 & 124$^a$ & 2.196$\pm$0.178 & 0.487$\pm$0.002 & -19.62 & 9.75 & 8.70 & 0.338 \\
93 & NGC 7743 & 0.57$^b$ & 1.03$^a$ & 3.60 & 84$^a$ & 1.058$\pm$0.103 & 0.293$\pm$0.001 & -20.18 & 9.82 & 16.97 & -- \\
94 & UGC 4551 & 0.51$^b$ & 0.82$^a$ & 2.89 & 167$^a$ & -0.156$\pm$0.002 & 0.145$\pm$0.001 & -19.78 & 9.70 & 3.10 & -- \\
95 & UGC 4587 & 0.81$^b$ & 1.05$^a$ & 3.81 & -- & -0.378$\pm$0.006 & 0.320$\pm$0.001 & -20.77 & -- & 4.77 & -- \\
96 & UGC 6062 & 0.82$^b$ & 1.01$^a$ & 3.26 & 142$^e$ & 0.339$\pm$0.005 & 0.250$\pm$0.001 & -20.34 & 9.93 & 3.53 & -- \\
97 & VCC 1199 & 0.90$^b$ & 0.54$^a$ & 3.97 & 55$^e$ & 0.484$\pm$0.015 & 0.025$\pm$0.001 & -15.58 & 9.82 & 10.00 & -- \\
98 & VCC 1440 & 0.89$^b$ & 0.54$^a$ & 2.41 & -- & 0.140$\pm$0.008 & 0.129$\pm$0.002 & -17.24 & -- & 3.30 & -- \\
99 & VCC 1545 & 0.51$^b$ & 0.54$^a$ & 3.30 & 51$^e$ & -0.215$\pm$0.012 & 0.144$\pm$0.004 & -17.49 & 9.08 & 2.89 & -- \\
100 & VCC 1627 & 0.69$^b$ & 0.54$^a$ & 1.93 & -- & 0.097$\pm$0.009 & 0.089$\pm$0.002 & -16.42 & -- & 1.78 & -- \\

\noalign{\smallskip}\hline
\end{longtable}
\tablecomments{\textwidth}{
See the notes in Table 1 for each column.
}

%%%%%%%%%%%%%%%%%%%%%%%%%%%%%%%%%%%%%%%%%%%%%%%%
\small
\setlength\LTleft{0pt}
\setlength\LTright{0pt}
\begin{longtable}{cccccccccccc}
\caption{``Intermediate'' Galaxy Parameters\label{Table3}}\\
  \hline\noalign{\smallskip}
N& Galaxy& $\gamma'$& $r_\gamma$& $R_{e}$& $\sigma$& $a_{4}/a$& $\varepsilon$& $M_V$ & $\log {M_{dyn}\over M_{\odot}}$ & $n$ & $\lambda_{R_e/2}$ \\
 &  & & $\log({\rm pc})$ & $\log({\rm pc})$ & ${\rm km \, \rm s^{-1}}$ & $10^{(-2)}$ & & ${\rm mag}$ & & & \\
(1) & (2) & (3) & (4) & (5) & (6) & (7) & (8) & (9) & (10) &(11) & (12) \\
  \hline\noalign{\smallskip}

1 & J091944.2+562201.1 & 0.32$^d$ & -- & 4.97 & 327$^e$ & -1.982$\pm$0.134 & 0.198$\pm$0.020 & -23.84 & 12.37 & 6.46 & -- \\
2 & J112842.0+043221.7 & 0.47$^d$ & -- & 4.01 & 360$^e$ & -1.068$\pm$0.194 & 0.249$\pm$0.021 & -22.86 & 11.49 & 3.09 & -- \\
3 & J120011.1+680924.8 & 0.33$^d$ & -- & 4.69 & 380$^e$ & 1.249$\pm$0.078 & 0.310$\pm$0.009 & -23.95 & 12.22 & 4.66 & -- \\
4 & J133724.7+033656.5 & 0.37$^d$ & -- & 4.83 & 414$^e$ & -1.755$\pm$0.079 & 0.167$\pm$0.008 & -22.50 & 12.43 & 10.05 & -- \\
5 & J135602.4+021044.6 & 0.40$^d$ & -- & 4.26 & 352$^e$ & -1.807$\pm$0.109 & 0.272$\pm$0.010 & -23.89 & 11.72 & 4.89 & -- \\
6 & J162332.4+450032.0 & 0.35$^d$ & -- & 4.65 & 356$^e$ & 0.142$\pm$0.068 & 0.202$\pm$0.007 & -23.14 & 12.12 & 4.82 & -- \\
7 & NGC 2841 & 0.34$^b$ & 1.09$^a$ & 3.76 & 206$^a$ & 1.778$\pm$0.001 & 0.286$\pm$0.001 & -20.57 & 10.75 & 12.98 & -- \\
8 & NGC 3998 & 0.49$^c$ & 0.83$^c$ & 2.71 & -- & 0.424$\pm$0.066 & 0.147$\pm$0.002 & -19.58 & -- & 2.17 & 0.342 \\
9 & NGC 4239 & 0.46$^b$ & 1.06$^a$ & 3.03 & 62$^a$ & 1.051$\pm$0.004 & 0.420$\pm$0.001 & -18.50 & 8.98 & 2.56 & -- \\
10 & NGC 4270 & 0.44$^c$ & 1.62$^c$ & 3.66 & -- & -0.527$\pm$0.048 & 0.420$\pm$0.001 & -20.79 & -- & 5.75 & 0.294 \\
11 & NGC 4350 & 0.47$^c$ & 1.57$^c$ & 3.24 & -- & 2.411$\pm$0.132 & 0.415$\pm$0.001 & -20.44 & -- & 5.00 & 0.480 \\
12 & NGC 4377 & 0.41$^c$ & 1.17$^c$ & 3.75 & -- & 0.587$\pm$0.035 & 0.157$\pm$0.001 & -19.90 & -- & 15.19 & 0.338 \\
13 & NGC 4379 & 0.46$^c$ & 1.02$^c$ & 3.18 & -- & 0.442$\pm$0.031 & 0.182$\pm$0.001 & -19.47 & -- & 5.04 & 0.300 \\
14 & NGC 4452 & 0.39$^c$ & 2.37$^c$ & 2.38 & 45$^e$ & 2.301$\pm$0.287 & 0.667$\pm$0.008 & -15.52 & 8.05 & 1.22 & 0.648 \\
15 & NGC 4476 & 0.34$^c$ & 2.32$^c$ & 3.25 & 49$^e$ & 2.432$\pm$0.175 & 0.426$\pm$0.003 & -20.27 & 9.01 & 2.92 & 0.266 \\
16 & NGC 4477 & 0.38$^c$ & 1.38$^c$ & 3.43 & -- & 2.203$\pm$0.069 & 0.218$\pm$0.001 & -21.05 & -- & 4.02 & 0.221 \\
17 & NGC 4482 & 0.49$^b$ & 2.05$^a$ & 3.59 & 26$^a$ & -0.260$\pm$0.005 & 0.324$\pm$0.001 & -18.87 & 8.79 & 2.40 & -- \\
18 & NGC 4733 & 0.35$^c$ & 1.90$^c$ & 3.55 & -- & 1.240$\pm$0.066 & 0.224$\pm$0.001 & -18.70 & -- & 4.60 & 0.076 \\
19 & NGC 4762 & 0.40$^c$ & 1.42$^c$ & 3.13 & -- & 2.335$\pm$0.056 & 0.513$\pm$0.002 & -20.20 & -- & 7.10 & 0.724 \\
20 & NGC 5422 & 0.45$^c$ & 1.30$^c$ & 3.97 & -- & 2.186$\pm$0.076 & 0.385$\pm$0.002 & -20.19 & -- & 12.22 & 0.501 \\
21 & NGC 5475 & 0.40$^c$ & 1.49$^c$ & 3.76 & 102$^e$ & 0.874$\pm$0.090 & 0.450$\pm$0.001 & -19.78 & 10.15 & 9.88 & 0.638 \\

\noalign{\smallskip}\hline
\end{longtable}
\tablecomments{\textwidth}{
See the notes in Table 1 for each column.
}
\end{center}

%%%%%%%%%%%%%%%%%%%%%%%%%%%%%%%%%%%%%%%%%%%%%%%%
\normalsize
\section{Results and Discussions}
\label{result}

\subsection{The relations of $a_{4}/a$ with central properties of ETGs}

\begin{figure}
 \centering
   \includegraphics[width=\textwidth, angle=0]{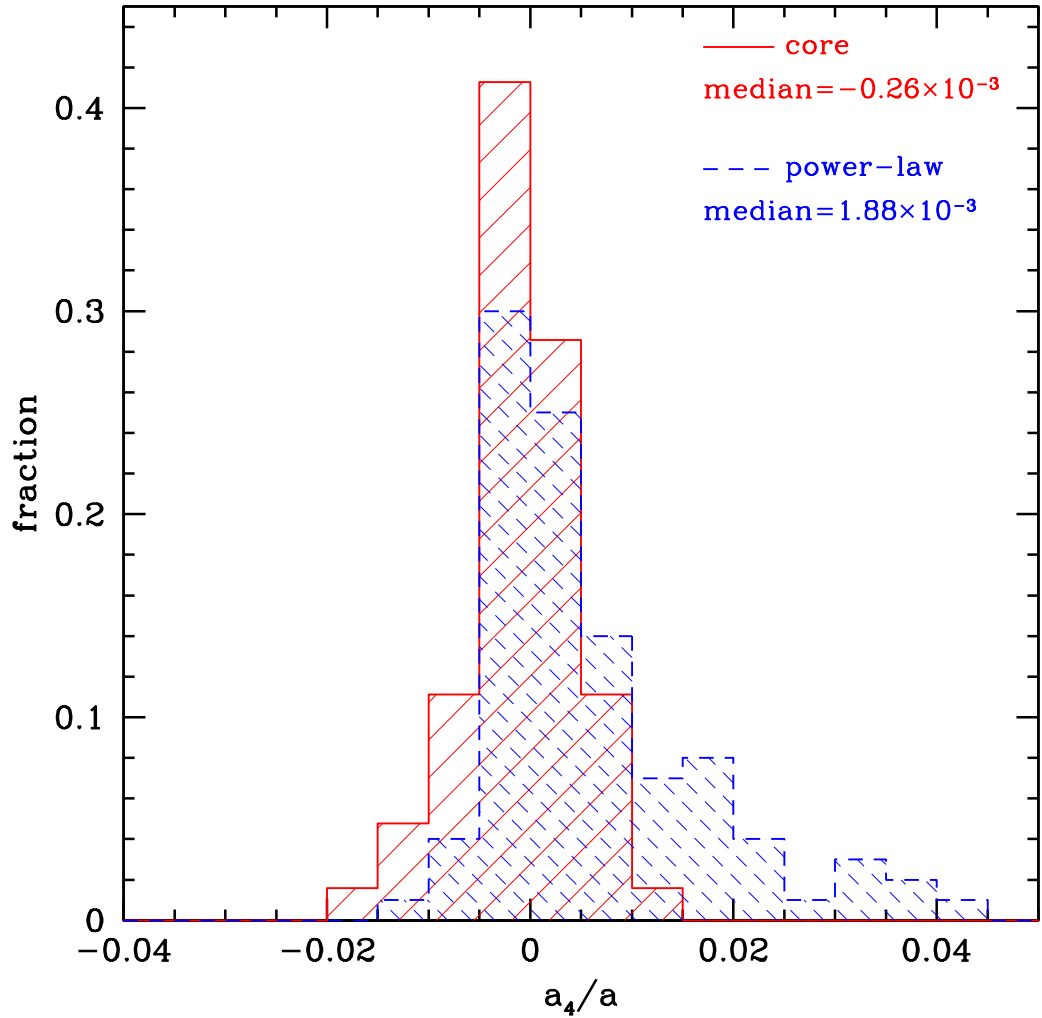}
\caption{Histograms of the isophotal shape characteristic parameter
$a_{4}/a$ for ``core'' (red solid line) and ``power-law'' (blue dashed line) ETGs.
The median values are indicated in the top right.
}
\label{Fig1}
\end{figure}

In this subsection, we re-visit the relations between the isophotal shape described
by characteristic parameter $a_{4}/a$ and the central photometric properties of ETGs.
Figure~\ref{Fig1} shows the distributions of $a_4/a$ for ``core'' and ``power-law'' ETGs.
It is clear from the histograms shown in Figure~\ref{Fig1} that the distribution
of $a_4/a$ for ``core'' and ``power-law'' ETGs is not significantly separated, i.e. 
the $a_4/a$ distribution of a large fraction of ``core'' and ``power-law''  ETGs
overlap. The fractions of boxy and disky galaxies are 59\% (37/63) and 41\% (26/63)
for ``core'' ETGs, and the median value of $a_4/a$ is $-0.26\times10^{-3}$.
While for the 100 ``power-law'' ETGs, the fractions of boxy and disky galaxies 
are $\sim$ 35\% (35/100) and 65\% (65/100), respectively, and $a_4/a$ has the median
value of $1.88\times10^{-3}$. It suggests that ``core'' galaxies are not necessarily
boxy, and only two-thirds ``power-law'' galaxies have disky deviated isophotes.
But the most deviated disky galaxies are ``power-law'' galaxies.

\begin{figure}
 \centering
   \includegraphics[width=\textwidth, angle=0]{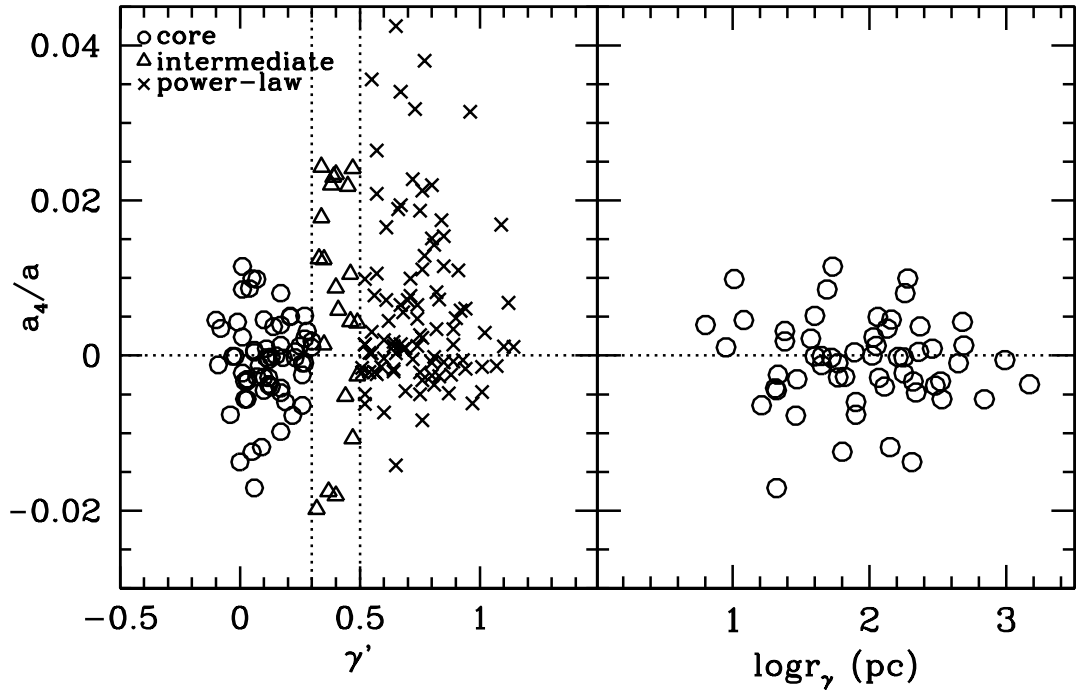}
\caption{Isophotal shape characteristic parameter $a_{4}/a$ as a function of
(a) central surface brightness profile slope $\gamma'$ (left panel) and 
(b) ``cusp radius'' $r_{\gamma}$ for ``core'' galaxies (right panel).
The horizontal dotted lines show the classification of disky ($a_{4}/a>0$)
and boxy ($a_{4}/a<0$) isophotal shape, while the vertical dotted lines in the 
left panel show the separation of ``core'' ($\gamma'<0.3$, circles), 
``power-law'' ($\gamma'>0.5$, crosses) and ``intermediate'' ETGs ($0.3<\gamma'<0.5$, triangles).
}
\label{Fig2}
\end{figure}

The left panel of Figure~\ref{Fig2} shows the $a_{4}/a$ as a function
of $\gamma'$ that describes the slope of the central surface brightness profile of
ETGs (see Lauer et al. 2007b). The two dotted vertical lines in
Figure~\ref{Fig2} divide sample ETGs into ``core'', ``intermediate'' and
``power-law'' galaxies, while the disky and boxy ETGs are located above and below
the horizontal dotted line. Although, there is a trend that the values of $a_{4}/a$
increase as $\gamma'$ increasing, the scatter shown in Figure~\ref{Fig2} is
quite large, i.e. the correlation between $a_{4}/a$ and $\gamma'$ is weak.
The Spearman rank-order correlation coefficient between $a_{4}/a$ and $\gamma'$
is $r_{s}=0.24$ and the probability that no correlation exists between these two
parameters is $1.12\times10^{-3}$. Apart from the central light profiles slope
$\gamma'$, the values of $r_{\gamma}$ characterize the physical scale of the core
of ETGs (``cusp radius'', Lauer et al. 2007a). Lauer et al. (2007a) claimed that the
core size $r_{\gamma}$ is tightly correlated with the galaxy luminosity $L$ and the
black hole mass $M_\bullet$. However, from the right panel of
Figure~\ref{Fig2} that shows the $a_{4}/a$ as a function of $r_{\gamma}$,
there is no correlation found between $a_{4}/a$ and $r_{\gamma}$ for ``core'' ETGs.

Note that Bender et al. (1989) pointed out that $a_{4}/a$ might be influenced by
the projection effect, but such effect could only lead to the change on its absolute
value, but not the sign of $a_{4}/a$. Thus the classification of boxy and disky ETGs
reflects the intrinsic isophotal property of ETGs. Therefore the statistical results
shown above do not support the statement that there is a close relation between
isophotal shape and their central light profile for ETGs.

\subsection{The relations of $a_{4}/a$ with global properties of ETGs}

\begin{figure}
 \centering
   \includegraphics[width=\textwidth, angle=0]{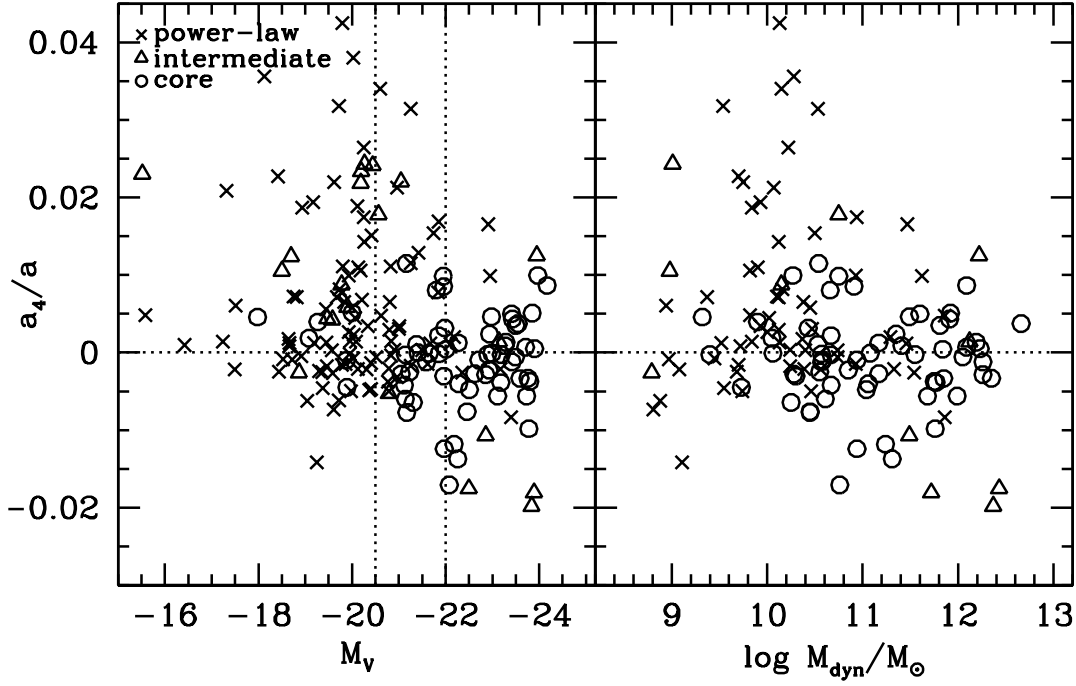}
\caption{
Isophotal shape parameter $a_{4}/a$ versus the $V$-band absolute magnitude $M_{V}$ (left panel) and the dynamical mass (right panel).
Circles represent ``core'' galaxies, triangles are ``intermediate'' galaxies,
and crosses show ``power-law'' galaxies.
The horizontal dotted line is the separation of disky ($a_{4}/a>0$) and boxy
($a_{4}/a<0$) ETGs.
The vertical dotted lines in the left panel indicate the less luminous galaxies 
with $M_V>-20.5$, luminous galaxies with $M_V<-22$ and ETGs in the interval 
of $-22<M_V<-20.5$ reported by Lauer et al. (2007b).
}
 \label{Fig3}
\end{figure}

In this subsection, we investigate the relations of the isophotal shape 
with other global physical properties for ``core'' and ``power-law'' ETGs. 
The left panel of Figure~\ref{Fig3} shows $a_{4}/a$ as a function of the 
$V$-band luminosity ($M_{V}$). It can be seen from the left panel of 
Figure~\ref{Fig3} that there is a weak correlation between $a_{4}/a$ and
$M_{V}$ for the whole sample ETGs, indicating that boxy (disky) ETGs tend to be 
bright (faint). 
Faber et al. (1997) pointed out that luminous galaxies with $M_V<-22$ mag have shallow
``core'' inner profiles, faint galaxies with $M_V>-20.5$ mag show steep ``power-law'' 
inner profiles, and for those with $-22<M_V<-20.5$, ``core'' and ``power-law''  
galaxies coexist. Therefore, it is interesting to visit the fractions of boxy and
disky galaxies in these luminosity intervals.
The fraction of boxy (disky) galaxies is $\sim$ 32\% (68\%) for faint ETGs with $M_{V}>-20.5$ mag,
while $\sim$ 58\% (42\%) for the most luminous ETGs with $M_{V}<-22$ mag.
For ETGs in the luminosity interval of $-22<M_{V}<-20.5$, the fractions of boxy and disky galaxies
are similar ($\sim$ 46\% and 54\%, respectively).
These results are consistent with Hao et al. (2006).
On the other hand, the fraction of ``core'' (``power-law'') is $\sim$ 7\% (79\%) for
faint ETGs with $M_{V}>-20.5$ mag, while $\sim$ 74\% (14\%) for the most luminous ETGs
with $M_{V}<-22$ mag. In the luminosity interval of $-22<M_{V}<-20.5$, the fractions are
comparable for ``core'' and ``power-law'' ETGs ($\sim$ 41\% and 53\%, respectively).
Thus we confirm the conclusion of Faber et al. (1997).
In the right panel of Figure~\ref{Fig3},
we show the isophotal shape parameter $a_{4}/a$ as a function of the dynamical mass.
Similar to the left panel of Figure~\ref{Fig3},
there is tendency that $a_{4}/a$ decreases as ETGs are more massive and ``core'' ETGs
are dominated by massive ETGs with dynamical mass larger than $10^{11} M_{\odot}$.
Therefore, the bimodal classification based on the central light profile (``core''
and ``power-law'') is more tightly correlated with galaxy luminosity and dynamical
mass than that based on isophotal shape (boxy and disky).

\begin{figure}
 \centering
   \includegraphics[width=\textwidth, angle=0]{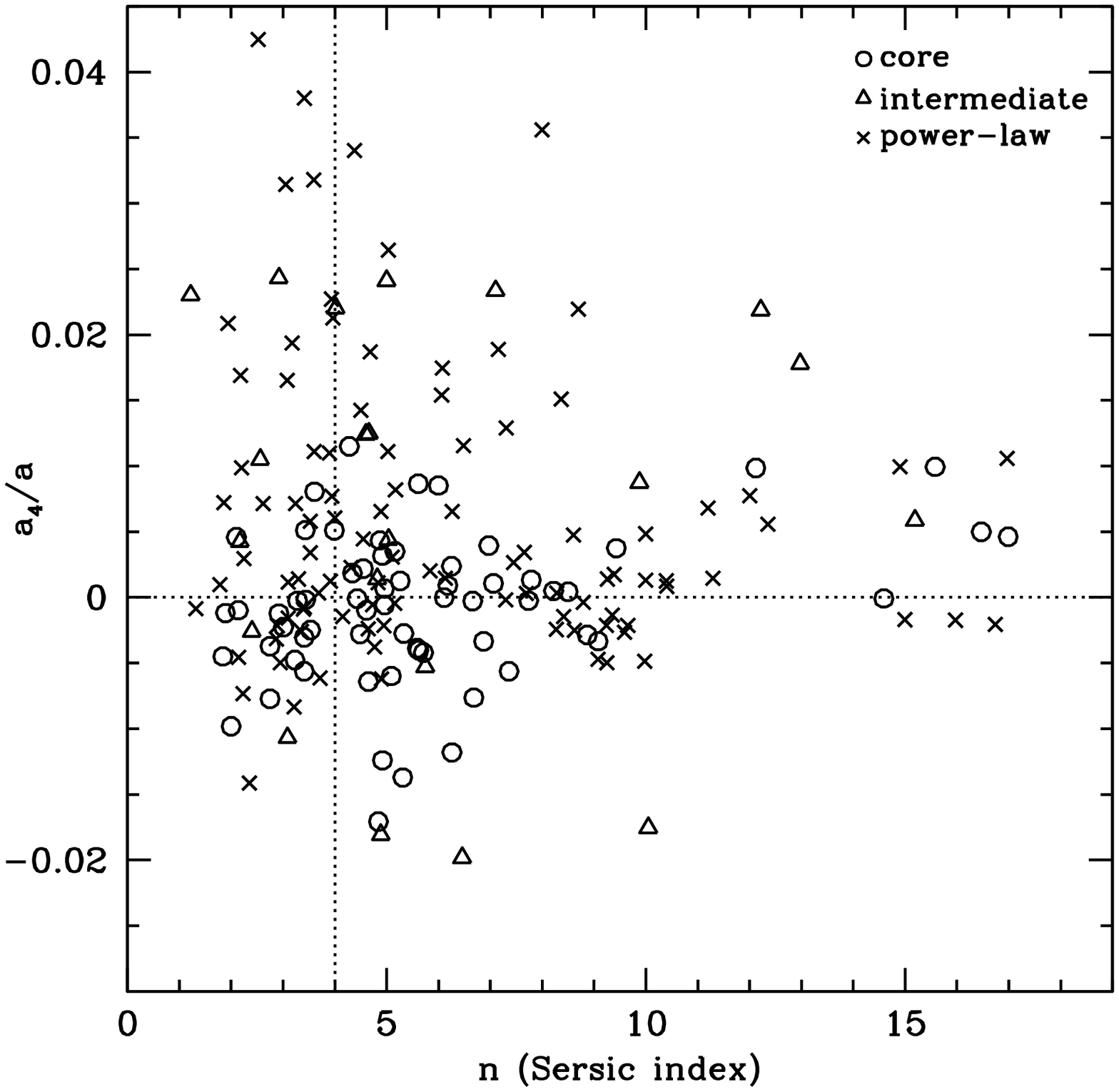}
\caption{
Isophotal shape parameter $a_{4}/a$ versus S\' ersic index n.
Circles represent ``core'' galaxies, triangles are ``intermediate'' galaxies,
and crosses show ``power-law'' galaxies.
The horizontal dotted line is the separation of disky ($a_{4}/a>0$) and boxy
($a_{4}/a<0$) galaxies.
The vertical dotted line indicates ETGs with $n=4$ following Kormendy (2009a).
}
 \label{Fig4}
\end{figure}

As is well known, the S\' ersic law is widely used to model the surface brightness profiles of
ETGs and the best fitting value of S\' ersic index $n$ could be used to describe the structures of galaxies. Based on photometric analysis for an ETGs sample in the Virgo cluster,
Kormendy (2009a) claimed that giant ETGs characterized by $n>4$ tend to be rotating slowly,
less flattened (ellipticity $\sim$ 0.15) and with boxy isophotes as well as with ``core''
in their center. To test whether this argument applies to our ETGs sample,
we plot $a_{4}/a$ as a function of the S\' ersic index $n$ in Figure~\ref{Fig4}.
It shows that there is no correlation between $a_{4}/a$ and the S\' ersic
index $n$ and the S\' ersic index $n$ is smaller than 4 for quite a few of ETGs with
``core'' or boxy isophotes, which indicates that the central and isophotal properties of ETGs
are not directly related to the S\' ersic index $n$.

\begin{figure}
   \centering
   \includegraphics[width=\textwidth, angle=0]{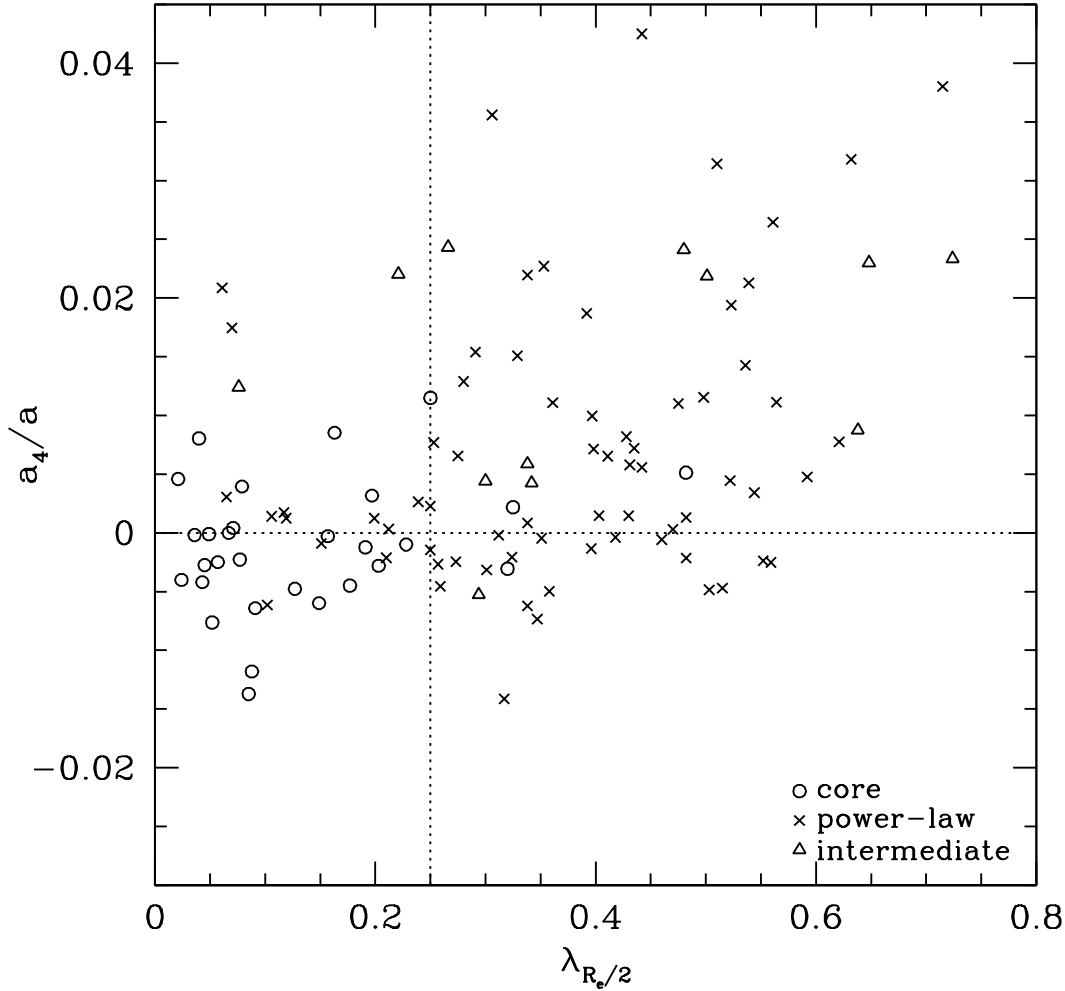}
   \caption{Isophotal shape parameter $a_{4}/a$ as a function of specific
angular momentum $\lambda_{R_e/2}$. 
The circles show ``core'' galaxies, triangles are ``intermediate'' galaxies, 
and crosses show ``power-law'' galaxies.
The horizontal dotted line separates disky ($a_{4}/a>0$) and boxy
($a_{4}/a<0$) galaxies.
The vertical dotted line shows the discriminator of $\lambda_{R_e/2}=0.25$ between slow and fast rotators.
}
\label{Fig5}
\end{figure}

Moreover, \atlas\ group (e.g. Cappellari et al. 2011) claimed that both isophotal shape and
central light profile of ETGs are secondary indicators of the galaxy kinematic
structure. They strongly suggest using the specific angular momentum parameter $\lambda_R$
as a discriminator of the bimodal distribution of ETGs, i.e. fast and slow rotators.
It is interesting to investigate relations among isophotal shapes, central light profiles and
the kinematic properties as parametrized by $\lambda_{R_e/2}$ that is the $\lambda_R$ measured
within half of the effective radius $R_e$. For our sample ETGs, the specific angular momentum
$\lambda_{R_e/2}$ is available for 111 objects from \atlas.
The sample size is 1.8 times that of Lauer (2012) who only compared the central light profile
to $\lambda_{R_e/2}$. Figure~\ref{Fig5} shows how $a_4/a$ varies with $\lambda_{R_e/2}$ for
``core'' and ``power-law'' ETGs. The dotted vertical line represents the $\lambda_{R_e/2}=0.25$
line dividing slow and fast rotators as suggested by Lauer (2012),
while the horizontal dotted line divides boxy and disky galaxies. It can be clearly
seen that there is a trend that $a_{4}/a$ increases as $\lambda_{R_e/2}$ increasing
and the ETGs with highest $a_4/a$ are fast rotators.
However, we can also see that such trend is only for power-law ETGs.
We note that $a_{4}/a$ and $\lambda_{R_e/2}$ are both affected by inclination effects.
But inclination does not change the classification by isophotal shape or $\lambda_{R_e/2}$
(Bender et al. 1989; Krajnovi\' c et al. 2013). So number statistics are more meaningful than
the trend. The fraction of disky (boxy) galaxies is $\sim$ 70\% (30\%) for fast rotators
($\lambda_{R_e/2}>0.25$), while $\sim$ 44\% (56\%) for slow rotators ($\lambda_{R_e/2}<0.25$).
Virtually Emsellem et al. (2011) already investigated the relation between isophotal shape
parameter $a_{4}/a$ and kinematics of ETGs as characterized by ${\lambda_R^N}_e = \lambda_{R_e} / \sqrt{\epsilon}$,
and concluded that there is not a simple correlation between these two parameters.
Note that ${\lambda_R^N}_e$ has been corrected for the inclination effects and ${\lambda_R^N}_e=0.31$
was used as a separator of slow and fast rotators by Emsellem et al. (2011).
We also examined the relation between $a_{4}/a$ and ${\lambda_R^N}_e$ based on our sample ETGs,
and found a similar result to Emsellem et al. (2011).
But interestingly the fractions of disky/boxy galaxies in fast/slow rotators, as classified by
${\lambda_R^N}_e=0.31$, are similar to those classified by
$\lambda_{R_e/2}=0.25$. Therefore, possible physical connection exists
between $a_{4}/a$ and kinematic properties of ETGs whether the characteristic angular momentum parameter
is influenced by projection effect.

\begin{figure}
   \centering
   \includegraphics[width=\textwidth, angle=0]{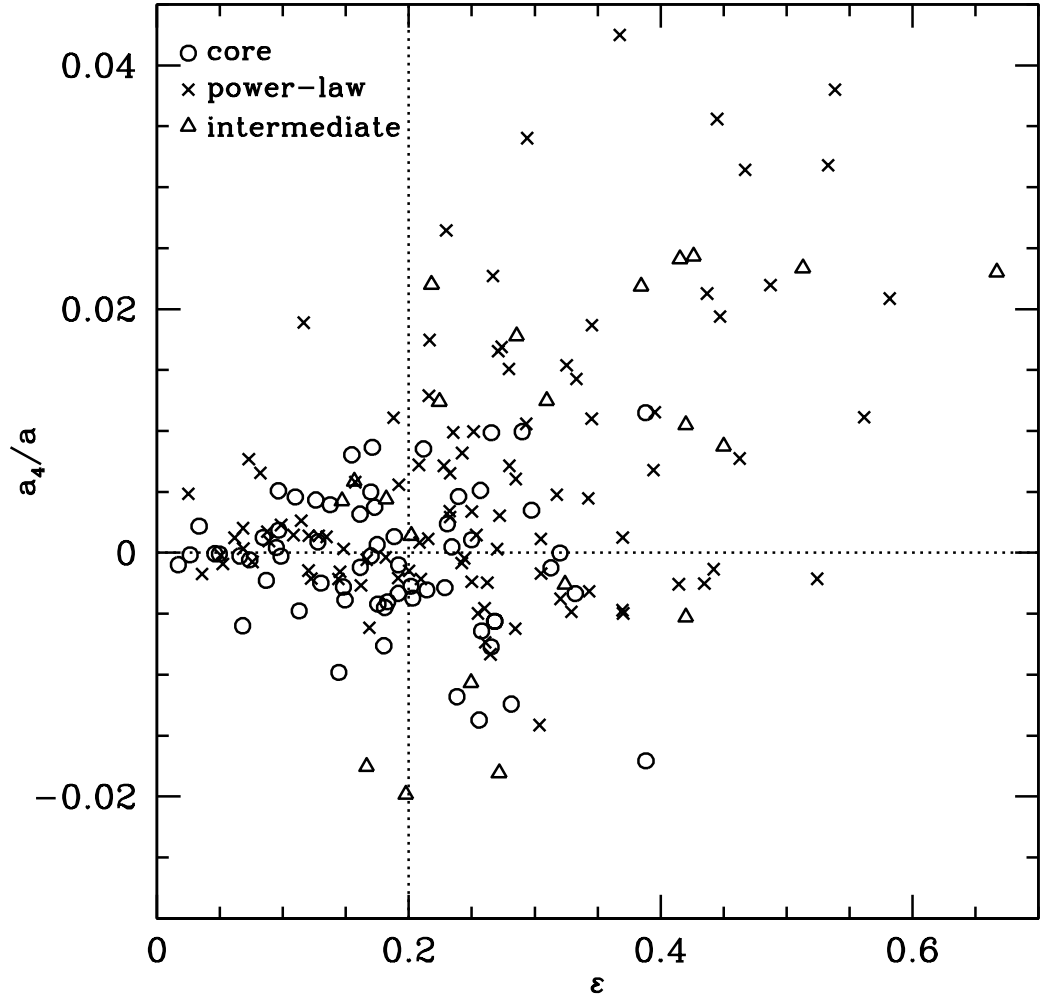}
   \caption{Isophotal shape parameter $a_{4}/a$ as a function of ellipticity $\epsilon$.
The red circles show ``core'' galaxies, triangles are ``intermediate'' galaxies,
and blue crosses show ``power-law'' galaxies.
The horizontal dotted line divides disky ($a_{4}/a>0$) and boxy
($a_{4}/a<0$) galaxies.
The vertical dotted line shows galaxies with ellipticity separator of $\epsilon=0.2$.
}
   \label{Fig6}
   \end{figure}

The \atlas\ group (e.g. Krajnovi\'c et al. 2013) and Lauer (2012) both found that
$\lambda_R$ is correlated with the ellipticity $\epsilon$, and the distribution of
``core'' and ``power-law''galaxies can be well separated in the $\lambda$-$\epsilon$ diagram.
It is interesting to study the relations among the isophotal shapes, ellipticity and nuclear profiles
of galaxies. Figure~\ref{Fig6} shows $a_4/a$ as a function of $\epsilon$ 
for ``core'' and ``power-law'' ETGs. It shows that there is a weak correlation between $a_{4}/a$ and
the ellipticity $\epsilon$ for ``power-law'' ETGs, 
and no correlation between these two parameters for ``core'' ETGs.
From Figure~\ref{Fig6}, we note that the most distorted disky galaxies have the largest 
ellipticity, which is coincident with that found by Hao et al. (2006). 
In that work, this was explained as a consequence of a biased viewing angle.
However, as shown in Figure~\ref{Fig1} and Figure~\ref{Fig5}, the ETGs with 
largest $a_4/a$ are also ``power-law'' and fast rotators. 
So orientation cannot account for the high value of $a_4/a$, which may be 
caused by some physical processes.
Specially, for those 111 ETGs that have $\lambda_{R_e/2}$ measurements available from \atlas,
the Spearman rank-order correlation coefficients and the probabilities that
no correlation exists for $\lambda_{R_e/2}$ versus $\epsilon$ and $a_4/a$ versus $\epsilon$
are $r_s=0.55, 0.38$ and Prob=$3.96\times10^{-10}$, $3.61\times10^{-5}$, respectively.
It indicates that the correlation between kinematic property and ellipticity is tighter than that between isophotal shape and ellipticity.

\section{Summary}
\label{sect:summary}

In this paper, we study the relations among isophotal
shapes, central light profiles and kinematic properties of ETGs based
on a compiled sample with 184 objects observed with both {\it HST} and SDSS DR8.
Our main results are summarized as follows:

\begin{enumerate}
\item There are no obvious relations of isophotal parameter $a_{4}/a$ with the central light
profile slope $\gamma'$ and the central ``cusp radius'' $r_{\gamma}$.
About 41\% ``core'' ETGs have disky isophotes, and 35\% ``power-law'' ETGs are boxy distorted.
\item Our statistical results show that there are only weak correlations between
$a_{4}/a$ and the galaxy luminosity $M_{V}$, and between $a_{4}/a$ and the dynamical mass.
Nuclear profiles correlate with $M_V$ and dynamical mass more tightly.
In addition, there is no any correlation between $a_{4}/a$ and the S\' ersic index $n$.
\item There are similar correlations between $a_{4}/a$ and ellipticity and
between $a_{4}/a$ and the specific angular momentum  $\lambda_{R_e/2}$, i.e.
$a_{4}/a$ is correlated with ellipticity and $\lambda_{R_e/2}$ for
``power-law'' ETGs, but no such relations for ``core'' ETGs.
Quite a large fraction of fast rotator ETGs (70\%) have disky isophotes,
while the slow rotator ETGs (56\%) tend to be boxy.
The most deviated disky galaxies (i.e. with highest $a_4/a$) are fast rotators 
and ``power-law'' ETGs. 
\end{enumerate}

Our statistical results support the statement by \atlas\ group
that isophotal shape ($a_{4}/a$) of ETGs has no simple relation with
both global and central properties of ETGs, but there seems to be correlation
between $a_{4}/a$ and kinematic property for ``power-law'' ETG. 
Considering that galaxy formation is a very complicated process as shown by both 
observations and simulations, which can lead to different morphologies, isophotal shapes,
central light profiles, kinematic and other global physical properties. There indeed
exist some trend among some physical parameters, but simple bimodal classifications
may be too simplistic.
As a caveat, our sample is compiled in a somewhat complicated way, 
so the numbers quoted in this paper may suffer from some selection effects.

\begin{acknowledgements}

We thank Drs. Shude Mao and C. G. Shu for advice and helpful discussions.
We also thank the anonymous referee for constructive comments. 
This project is supported by the NSF of China 10973011, 10833006, 11003015, the
Chinese Academy of Sciences and NAOC (SM).  The Project-sponsored by SRF for
ROCS, SEM.  Funding for the creation and distribution of the SDSS Archive has
been provided by the Alfred P. Sloan Foundation, the Participating
Institutions, the National Aeronautics and Space Administration, the National
Science Foundation, the U.S. Department of Energy, the Japanese Monbukagakusho,
and the Max Planck Society. The SDSS Web site is http://www.sdss3.org/.  The
SDSS is managed by the Astrophysical Research Consortium (ARC) for the
Participating Institutions. The Participating Institutions are The University
of Chicago, Fermilab, the Institute for Advanced Study, the Japan Participation
Group, The Johns Hopkins University, the Korean Scientist Group, Los Alamos
National Laboratory, the Max-Planck-Institute for Astronomy (MPIA), the
Max-Planck-Institute for Astrophysics (MPA), New Mexico State University,
University of Pittsburgh, Princeton University, the United States Naval
Observatory, and the University of Washington.

\end{acknowledgements}

%\appendix                  

\clearpage

\label{lastpage}

\end{document}